\documentclass[twocolumn,tighten,times]{aastex63}
\usepackage{graphicx}

\newcommand{\src}{G350.1$-$0.3}

\catcode`\@=11
\newcommand{\gapprox}{\mathrel{\mathpalette\@versim>}}
\newcommand{\lapprox}{\mathrel{\mathpalette\@versim<}}
\newcommand{\propapprox}{\mathrel{\mathpalette\@versim\propto}}
\newcommand{\@versim}[2]
  {\lower3.1truept\vbox{\baselineskip0pt\lineskip0.5truept
\ialign{$\m@th#1\hfil##\hfil$\crcr#2\crcr\sim\crcr}}}
\catcode`\@=12

\received{2020 November 11}
\revised{2020 November 24}
\accepted{2020 November 25}
\journalinfo{Astrophysical Journal Letters}

\shorttitle{EXPANSION OF G350.1$-$0.3}

\begin{document}

\title{Expansion and Age of the Supernova Remnant
  G350.1$-$0.3:  High-Velocity Iron Ejecta from a Core-Collapse Event}


\author[0000-0002-2614-1106]{Kazimierz J. Borkowski}
\affiliation{Department of Physics, North Carolina State University, 
Raleigh, NC 27695-8202, USA}

\author[0000-0001-6212-287X]{William Miltich}
\affiliation{Department of Physics, North Carolina State University, 
Raleigh, NC 27695-8202, USA}

\author[0000-0002-5365-5444]{Stephen P. Reynolds}
\affiliation{Department of Physics, North Carolina State University, 
Raleigh, NC 27695-8202, USA}

\begin{abstract}

  We report {Chandra} observations of the highly asymmetric
  core-collapse supernova remnant \src.  We document expansion over 9
  yr away from the roughly stationary central compact object, with
  sky-plane velocities up to $5000\, d_{4.5}$ km s$^{-1}$ ($d_{4.5}$ is the
  distance in units of 4.5 kpc),
  redshifts ranging from $900$ km s$^{-1}$ to 2600 km s$^{-1}$, and
  three-dimensional space velocities approaching 6000 km s$^{-1}$.
  Most of the bright emission comes from heavy-element ejecta
  particularly strong in iron.  Iron-enhanced ejecta are seen at
  4000--6000 km s$^{-1}$, strongly suggesting that the supernova was not
  a common Type IIP event.  While some fainter regions have roughly
  solar abundances, we cannot identify clear blast-wave features.  Our
  expansion proper motions indicate that \src\ is no more than about
  600 yr old, independent of distance: the third youngest known
  core-collapse supernova in the Galaxy, and one of the most
  asymmetric.
  
\end{abstract}

\keywords{
Supernova remnants (1667);
Core-collapse supernovae (304);
Ejecta (453);
X-ray astronomy (1810);
}

\section{Introduction}
\label{intro}

Young core-collapse supernova remnants (CC SNRs) can provide essential
information on the progenitors, immediate environments, and explosion
mechanisms of massive stars.  However, young CC SNRs are surprisingly
scarce in the Galaxy.  From the past millennium we know of four very
disparate objects: Cas A, age $\sim 350$ yr \citep[e.g.,][]{thor01};
Kes 75 (G29.7$-0.3$), $\sim 500$ yr \citep{reynolds18}; the Crab
Nebula, 966 yr; and G330.2$+1.0$, $\sim 1000$ yr \citep{borkowski18}.
Only Cas A has the
appearance one might expect of a somewhat spherical explosion in a
somewhat uniform ambient medium.  Kes 75 shows only half a shell
morphology, and the Crab none, and G330.2+1.0 has an extremely faint,
asymmetric shell.  Either the explosions themselves or their immediate
surroundings, or both, are highly irregular.

Evidence continues to accumulate for the intrinsic asymmetry of the CC
events themselves.  Asymmetries formed in the first few hundred
milliseconds can persist to ages of years
\citep[e.g.,][]{wongwathanarat17, gabler20, orlando20}.  Remnants
interact with their surroundings as well, giving clues to the
immediate pre-explosion environment.  One indicator of explosion
asymmetry is the evidence in a few SNe and SNRs of overturn in the
ejecta, that is, of high-velocity iron-group elements (IGEs) in
SNe, or of their presence at large radii in SNRs.
A well-known example of the former is SN 1987A, where detection of
prompt gamma-rays as well as direct measure of Fe line profiles
demanded the mixing of IGEs to 3000 km s$^{-1}$
\citep[e.g.,][]{mccray16}.  In SN 1993J, modeling of the
nebular spectrum implied the mixing of Fe out to at least 3000 km
s$^{-1}$\citep{houck96}.  Several other supernovae have shown evidence
for Fe at velocities between 6500 and 8400 km s$^{-1}$ \citep[][and
  references therein]{utrobin19}.  High IGE velocities are also
found in Cas A, up to 4500 km s$^{-1}$ \citep{delaney10}.
It is not known how widespread this phenomenon is among CC SNRs; its
demonstration in more cases will provide both a spur and a constraint
to modelers.

\begin{figure}
\centerline{\includegraphics[width=3.5truein]{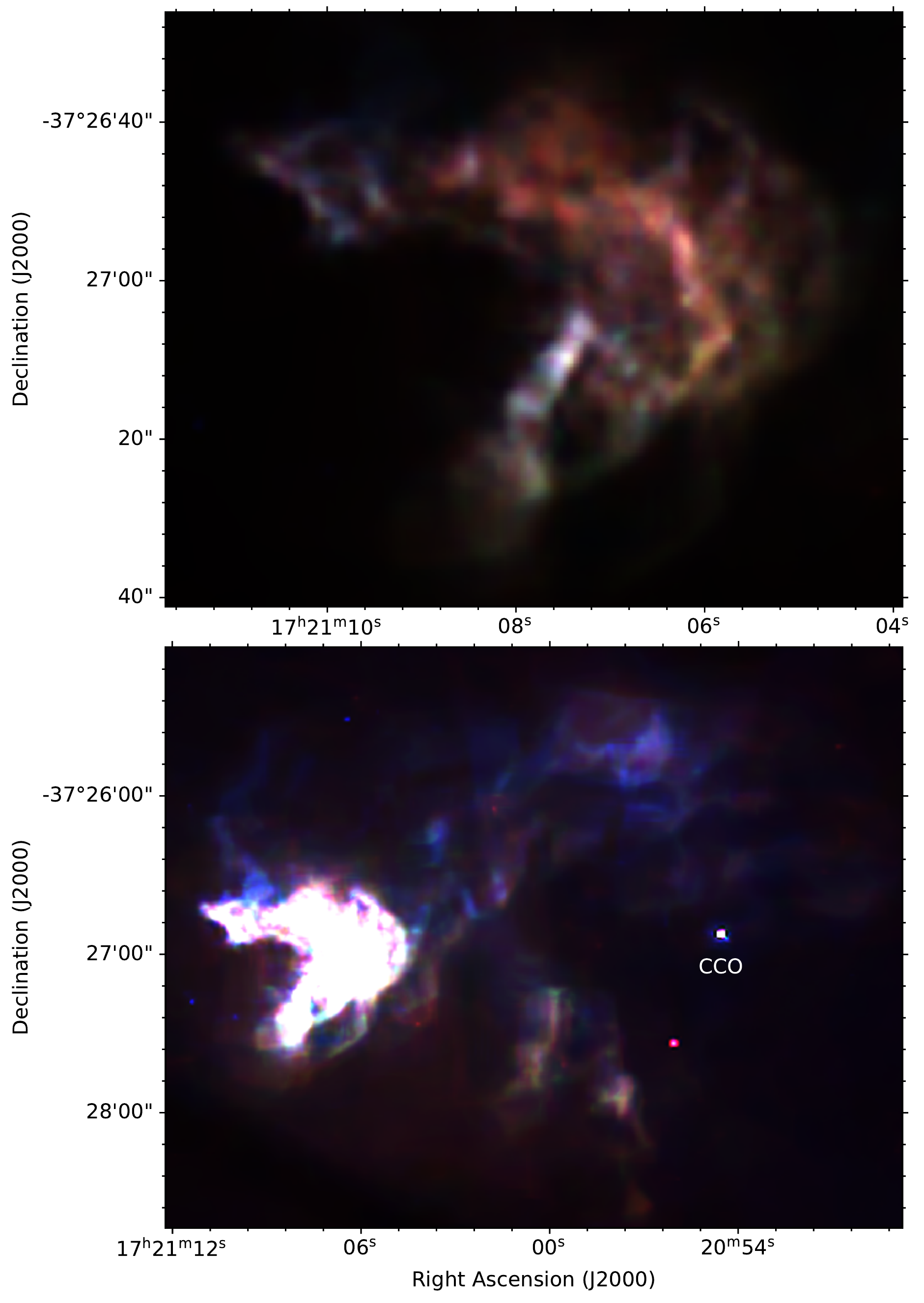}}
\caption{\small{Three-color {Chandra} image of \src, from 2018.  Red,
    0.5--1.6 keV; green, 1.6--2.6 keV; blue, 2.6--7 keV.
    Top:  close-up of the bright eastern region.  Bottom:  entire remnant.
    Note the strong spectral variations.}
}    
\label{3colors}
\end{figure}

Here we confirm a fifth member to add to the list of CC SNe of the
past millennium: \src, by far the most asymmetric of all
(Figure~\ref{3colors}).  Discovered in radio \citep{salter86}, its
identification as a single SNR had to wait for X-ray observations
\citep{gaensler08}, which also identified a nearby point X-ray source,
XMMU J172054.5-372652, and argued for its association with the
remnant, though pulsations were not detected.  This point source is
thus presumed to be a compact central object (CCO) as seen in several
other young SNRs. Those and subsequent observations
\citep[][]{lovchinsky11} led to an age estimate of 600--1200 yr,
based on simple spectral models.  Spectra showed strong overabundances
of Mg, Si, S, Ar, Ca, and Fe; emission appeared to be dominated by SN
ejecta in all areas.  The distance is still uncertain;
\cite{gaensler08} quote 4.5 kpc based on an inferred high preshock
density of 25 cm$^{-3}$ and therefore an association with a nearby
molecular cloud at that distance, while \cite{yasumi14} used 
Suzaku observations and different methods to obtain $9 \pm 3$ kpc.
We shall quote results in terms of $d_{4.5}$, the distance in units of
4.5 kpc.  \cite{yasumi14} reported a significant anomaly in their
analysis of the spatially integrated spectrum of \src: an
overabundance of stable nickel relative to iron: Ni/Fe $= 12 \pm 7$
times the solar value (mass ratio of $0.7 \pm 0.4$).  The combination
of a dramatically asymmetric morphology and this nickel excess
makes \src\ a particularly interesting object for detailed study.

\section{{Chandra} Observations}
\label{obssec}

\src\ was observed in 2009 May (PI: P.~Slane), and we observed it in
five segments in 2018 July for a total of 189 ks.
The mean separation of the two epochs is
9.126 yr.  In all observations, \src\ was placed on the Advanced CCD
Imaging Spectrometer (ACIS) S3 chip, with Very Faint mode used to
reduce the particle background everywhere except for the bright east region
where Faint mode was employed.

We aligned individual 2018 pointings using the CCO, while the inter-epoch alignment
was done by matching the positions of seven point sources near the {Chandra}
optical axis. We jointly fit them with 2D Gaussians to obtain best estimates of
image shifts between epochs, with statistical $1\sigma$ 
errors in alignment not exceeding $0.1$ ACIS pixels in R.A.~and decl.~(the ACIS pixel
size is $0\farcs492 \times 0\farcs492$). These errors translate into proper motion 
(PM) errors in R.A.~and decl.~not exceeding 5 mas yr$^{-1}$. We find a modest
(0.27 ACIS pixels) apparent displacement of the CCO toward the north--northwest,
which 
might be due to
a PM with $\mu_{\alpha}\cos \delta = -5$ mas yr$^{-1}$
and $\mu_{\delta} = 14$ mas yr$^{-1}$ ($v_t = 320 d_{4.5}$ km s$^{-1}$).
Both the magnitude and direction of the CCO motion are 
quite uncertain due to the large PM errors.

After alignment, we extracted images and spectra from event files. The
individual 2018 event files were merged together prior to image
extraction, but spectra were extracted separately from each individual
event file. We then summed them together and averaged their spectral
and ancillary responses by weighting them by individual exposure
times.

The combined 2018 image is shown in Figure~\ref{3colors}, where red
corresponds to 0.5--1.6 keV, green to 1.6--2.6 keV, and blue to 2.6--7
keV.  Substantial spectral variations are apparent and will be
discussed below.

\begin{figure*}
  \centerline{\includegraphics[width=6.5truein]{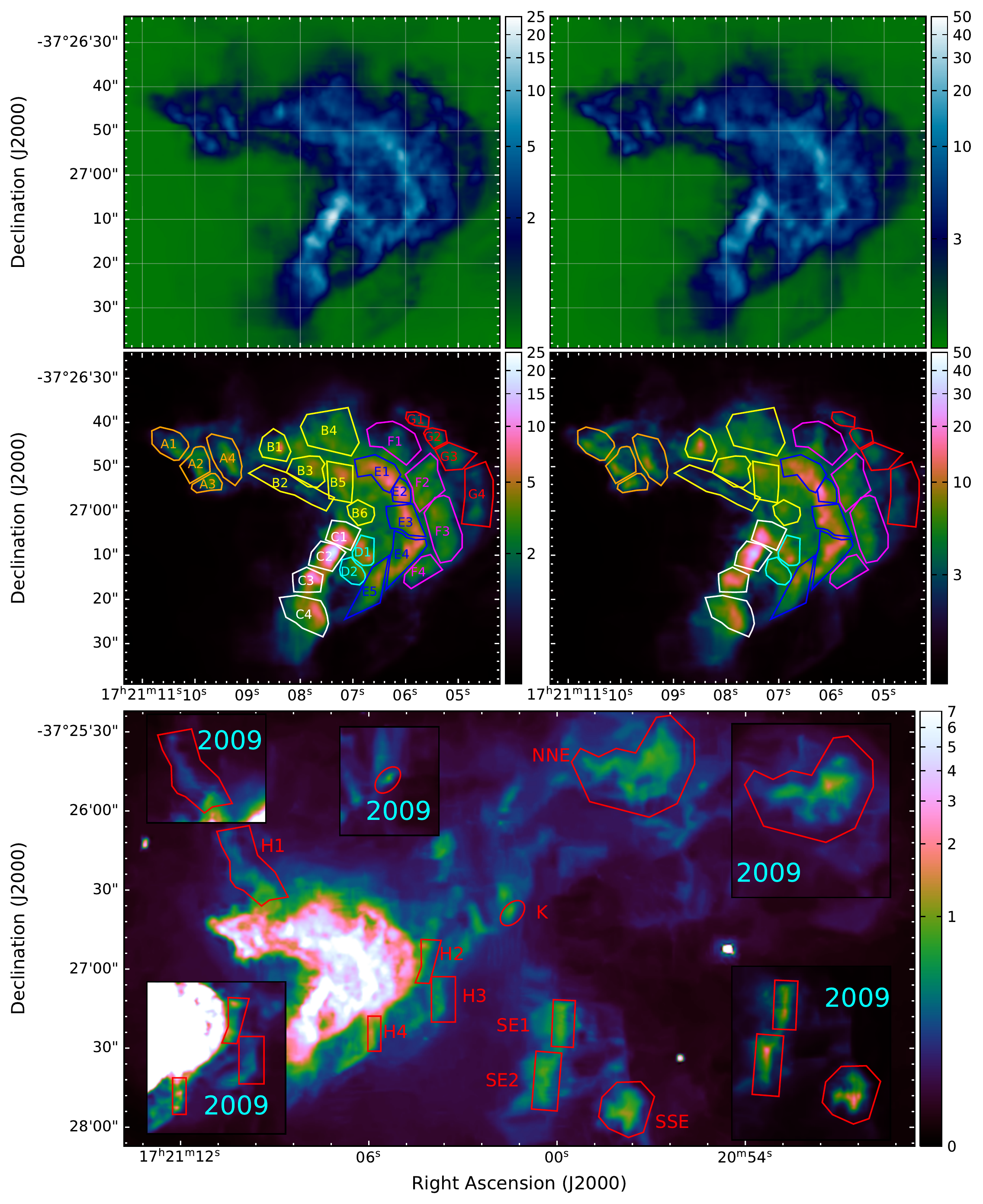}}
  \caption{Comparison of broadband  (0.5--7 keV) images at 
    the two epochs.  Top two rows:  (left) 2009; (right) 2018.  Regions used for
    measuring expansion are shown in the second row.  Bottom:  2018 image with local
    insets showing corresponding 2009 images for selected regions. Note
  the changes in morphology as well. The scale is in counts per
    $0 \farcs 296 \times 0 \farcs 296$ image pixel.}
  \label{regions}
    \end{figure*}

\section{Expansion}

Expansion between 2009 and 2018 is apparent by eye over most of the
remnant.  See Figure~\ref{regions}, which compares the 2009 (left) and
2018 (right) images of the bright E region in the top two rows, and displays
insets from the 2009 image on a 2018 image in the bottom.  We measured
expansion in the discrete regions (labeled in Figure~\ref{regions})
using the same maximum-likelihood method we used for RCW 89
\citep{borkowski20}; we smoothed the 2018 0.5--7 keV image to
use as a model, and
fit the unsmoothed 2009 0.5--7 keV data to it, calculating shifts, and
errors, in two directions.  This method sufficed for the
bright regions A--G; in fainter ones we used a Markov chain Monte
Carlo (MCMC) method for the fitting, as described in
\cite{borkowski18}. But for the faintest knot K,
we fit a 2D Gaussian (+ constant
background) to unsmoothed data at both epochs, then
calculated shifts between Gaussian peaks and converted
them into PMs.

\begin{figure*}
  \centerline{\includegraphics[width=6.0truein]{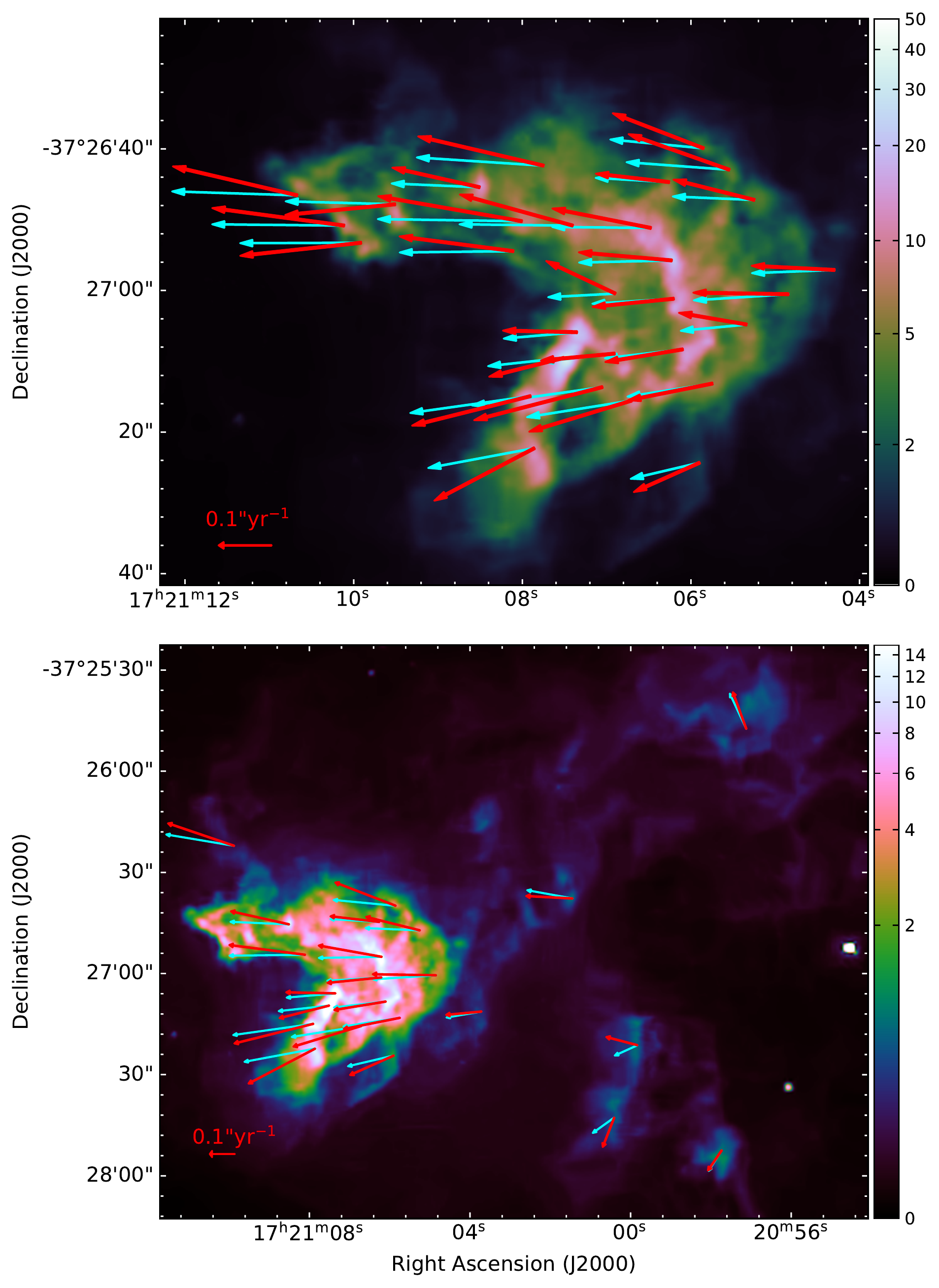}}
  \caption{Proper motions of expansion.  In both panels, total proper
    motions are shown in red, and radial components (i.e., directly
    away from the CCO) are in cyan.  Nonradial motions are nonnegligible.
    In the top panel, motions diverge from radial around the unseen
    obstacle creating the indentation. The scale is in counts per
    $0 \farcs 296 \times 0 \farcs 296$ image pixel.}
  \label{motions}
  \end{figure*}

Table~\ref{rates} gives our measured PMs (arcsec
yr$^{-1}$) in R.A.~and decl., with errors (in mas, in parentheses).
We also report radial (away from the CCO) and tangential
motions.
They are
shown in Figure~\ref{motions}, where cyan arrows show the radial
component of PM, while red arrows show the total.  Some
substantially nonradial motions are apparent.  Table~\ref{rates}
also gives expansion ages (distances to the CCO divided by radial
PMs).

\begin{deluxetable*}{ccllllcrcrl}   
  \tabletypesize{\footnotesize}
\tablecolumns{11}
\tablecaption{Proper Motions and Expansion Rates\label{rates}}
\tablehead{\colhead{Region} & \colhead{$R$} & \colhead{$\mu_{\alpha} \cos \delta$\tablenotemark{a}} & \colhead{$\mu_{\delta}$\tablenotemark{b}} & \colhead{$\mu_{r}$\tablenotemark{c}} & \colhead{$\mu_{t}$\tablenotemark{d}}  & \colhead{$v_{r}$\tablenotemark{e}} & \multicolumn{1}{c}{$v_{t}$\tablenotemark{e}} & \colhead{Expansion} & \multicolumn{1}{c}{$t_{f}$\tablenotemark{f}} & \colhead{$m$\tablenotemark{g}} \\
  & \colhead{($\arcsec$)} & \colhead{($\arcsec$ yr$^{-1}$)} & \colhead{($\arcsec$ yr$^{-1}$)} & \colhead{($\arcsec$ yr$^{-1}$)} & \colhead{($\arcsec$ yr$^{-1}$)} & \colhead{(km s$^{-1}$)} & \multicolumn{1}{c}{(km s$^{-1}$)} &\colhead{(\% yr$^{-1}$)} & \multicolumn{1}{c}{(yr)}}
\startdata
A1 & 192 & $0.235(28)$ & $-0.055(14)$ & $0.236(28)$ & $-0.047(13)$ & $5040 \pm 600$ & $-1010 \pm 280$ & $0.123 \pm 0.015$ & 810 & 0.74 \\
A2 & 185 & $0.249(7)$ & $-0.033(10)$ & $0.249(7)$ & $-0.031(10)$ & $5320 \pm 160$ & $-650 \pm 210$ & $0.135 \pm 0.004$ & 740 & 0.81 \\
A3 & 183 & $0.228(13)$ & \phantom{$-$}$0.025(11)$ & $0.228(13)$ & \phantom{$-$}$0.024(11)$ & $4870 \pm 270$ & \phantom{$-$}$510 \pm 230$ & $0.125 \pm 0.007$ & 800 & 0.75 \\
A4 & 178 & $0.208(8)$ & \phantom{$-$}$0.019(12)$ & $0.207(8)$ & \phantom{$-$}$0.026(12)$ & $4410 \pm 180$ & \phantom{$-$}$550 \pm 250$ & $0.116 \pm 0.005$ & 860 & 0.70 \\
B1 & 166 & $0.165(7)$ & $-0.037(9)$ & $0.166(7)$ & $-0.030(9)$ & $3550 \pm 140$ & $-640 \pm 200$ & $0.100 \pm 0.004$ & 1000 & 0.60 \\
B2 & 163 & $0.215(16)$ & $-0.028(10)$ & $0.215(16)$ & $-0.030(10)$ & $4580 \pm 340$ & $-650 \pm 210$ & $0.132 \pm 0.010$ & 760 & 0.79 \\
B3 & 160 & $0.271(13)$ & $-0.047(9)$ & $0.272(13)$ & $-0.043(8)$ & $5800 \pm 280$ & $-920 \pm 180$ & $0.170 \pm 0.008$ & 590 & 1.02 \\
B4 & 157 & $0.237(9)$ & $-0.055(12)$ & $0.240(9)$ & $-0.040(12)$ & $5120 \pm 190$ & $-850 \pm 260$ & $0.153 \pm 0.006$ & 660 & 0.91 \\
B5 & 153 & $0.213(12)$ & $-0.059(11)$ & $0.214(12)$ & $-0.056(11)$ & $4560 \pm 260$ & $-1200 \pm 240$ & $0.140 \pm 0.008$ & 720 & 0.84 \\
B6 & 147 & $0.131(13)$ & $-0.062(19)$ & $0.127(13)$ & $-0.068(19)$ & $2710 \pm 280$ & $-1460 \pm 400$ & $0.086 \pm 0.009$ & 1160 & 0.52 \\
C1 & 153 & $0.140(5)$ & $-0.003(6)$ & $0.139(5)$ & $-0.015(6)$ & $2970 \pm 100$ & $-320 \pm 130$ & $0.091 \pm 0.003$ & 1100 & 0.55 \\
C2 & 155 & $0.141(4)$ & \phantom{$-$}$0.036(6)$ & $0.144(5)$ & \phantom{$-$}$0.020(6)$ & $3080 \pm 100$ & \phantom{$-$}$430 \pm 120$ & $0.093 \pm 0.003$ & 1070 & 0.56 \\
C3 & 160 & $0.224(6)$ & \phantom{$-$}$0.056(7)$ & $0.229(6)$ & \phantom{$-$}$0.024(7)$ & $4890 \pm 130$ & \phantom{$-$}$510 \pm 140$ & $0.143 \pm 0.004$ & 700 & 0.86 \\
C4 & 161 & $0.189(7)$ & \phantom{$-$}$0.099(9)$ & $0.204(6)$ & \phantom{$-$}$0.063(9)$ & $4350 \pm 140$ & \phantom{$-$}$1340 \pm 190$ & $0.127 \pm 0.004$ & 790 & 0.76 \\
D1 & 148 & $0.140(9)$ & \phantom{$-$}$0.013(9)$ & $0.140(9)$ & $-0.002(9)$ & $2990 \pm 200$ & $-50 \pm 190$ & $0.095 \pm 0.006$ & 1060 & 0.57 \\
D2 & 150 & $0.242(12)$ & \phantom{$-$}$0.062(14)$ & $0.248(12)$ & \phantom{$-$}$0.028(14)$ & $5300 \pm 260$ & \phantom{$-$}$600 \pm 290$ & $0.165 \pm 0.008$ & 600 & 0.99 \\
E1 & 142 & $0.186(11)$ & \phantom{$-$}$0.037(8)$ & $0.186(11)$ & $-0.035(8)$ & $3970 \pm 240$ & $-740 \pm 180$ & $0.131 \pm 0.008$ & 760 & 0.79 \\
E2 & 139 & $0.176(6)$ & \phantom{$-$}$0.015(11)$ & $0.175(6)$ & $-0.019(11)$ & $3740 \pm 130$ & $-410 \pm 240$ & $0.126 \pm 0.005$ & 790 & 0.76 \\
E3 & 139 & $0.154(7)$ & \phantom{$-$}$0.015(11)$ & $0.154(7)$ & \phantom{$-$}$0.006(11)$ & $3290 \pm 150$ & \phantom{$-$}$130 \pm 240$ & $0.111 \pm 0.005$ & 900 & 0.67 \\
E4 & 138 & $0.147(7)$ & \phantom{$-$}$0.024(8)$ & $0.148(8)$ & \phantom{$-$}$0.007(7)$ & $3160 \pm 160$ & \phantom{$-$}$150 \pm 150$ & $0.107 \pm 0.006$ & 930 & 0.64 \\
E5 & 146 & $0.193(11)$ & \phantom{$-$}$0.059(16)$ & $0.200(13)$ & \phantom{$-$}$0.028(14)$ & $4260 \pm 280$ & \phantom{$-$}$590 \pm 300$ & $0.137 \pm 0.009$ & 730 & 0.82 \\
F1 & 140 & $0.101(10)$ & $-0.097(14)$ & $0.139(9)$ & $-0.008(14)$ & $2980 \pm 200$ & $-170 \pm 310$ & $0.100 \pm 0.007$ & 1000 & 0.60 \\
F2 & 134 & $0.151(10)$ & $-0.035(11)$ & $0.150(10)$ & $-0.040(10)$ & $3200 \pm 210$ & $-850 \pm 220$ & $0.112 \pm 0.007$ & 890 & 0.67 \\
F3 & 128 & $0.118(9)$ & \phantom{$-$}$0.016(15)$ & $0.121(9)$ & \phantom{$-$}$0.005(14)$ & $2590 \pm 190$ & \phantom{$-$}$100 \pm 310$ & $0.095 \pm 0.007$ & 1060 & 0.57 \\
F4 & 135 & $0.160(20)$ & \phantom{$-$}$0.032(13)$ & $0.163(20)$ & \phantom{$-$}$0.007(13)$ & $3470 \pm 420$ & \phantom{$-$}$150 \pm 280$ & $0.121 \pm 0.015$ & 830 & 0.72 \\
G1 & 135 & $0.172(15)$ & $-0.066(13)$ & $0.177(16)$ & $-0.050(13)$ & $3780 \pm 330$ & $-1060 \pm 270$ & $0.131 \pm 0.012$ & 760 & 0.79 \\
G2 & 131 & $0.190(11)$ & $-0.067(19)$ & $0.195(11)$ & $-0.053(19)$ & $4150 \pm 240$ & $-1130 \pm 400$ & $0.148 \pm 0.008$ & 670 & 0.89 \\
G3 & 127 & $0.153(14)$ & $-0.038(13)$ & $0.155(14)$ & $-0.032(13)$ & $3300 \pm 300$ & $-680 \pm 280$ & $0.121 \pm 0.011$ & 820 & 0.73 \\
G4 & 120 & $0.146(17)$ & \phantom{$-$}$0.002(21)$ & $0.146(16)$ & $-0.002(21)$ & $3120 \pm 330$ & $-40 \pm 440$ & $0.122 \pm 0.013$ & 820 & 0.73 \\
H1 & 185 & $0.188^{+18}_{-18}$ & \phantom{$-$}$0.064^{+20}_{-21}$ & $0.195^{+19}_{-19}$ & $-$0.032$^{+20}_{-20}$ & $4170^{+400}_{-400}$ & $-680^{+430}_{-430}$ & $0.106^{+0.010}_{-0.010}$ & 950 & 0.63 \\
H2 & 116 & $0.157^{+22}_{-22}$ & \phantom{$-$}$0.008^{+20}_{-19}$ & $0.156^{+22}_{-22}$ & $-$0.014$^{+19}_{-20}$ & $3340^{+460}_{-460}$ & $-300^{+410}_{-420}$ & $0.135^{+0.019}_{-0.019}$ & 740 & 0.81 \\
H3 & 111 & $0.101^{+26}_{-28}$ & $-0.010^{+38}_{-37}$ & $0.101^{+27}_{-28}$ & $-0.008^{+36}_{-38}$ & $2160^{+570}_{-590}$ & $-160^{+760}_{-800}$ & $0.091^{+0.024}_{-0.025}$ & 1090 & 0.55 \\
H4 & 139 & $0.124^{+16}_{-16}$ & $-0.055^{+23}_{-24}$ & $0.134^{+18}_{-18}$ & \phantom{$-$}$0.026^{+23}_{-22}$ & $2850^{+390}_{-380}$ & \phantom{$-$}$550^{+500}_{-480}$ & $0.096^{+0.013}_{-0.013}$ & 1040 & 0.58 \\
K & 83 & $0.132^{+37}_{-37}$ & \phantom{$-$}$0.007^{+39}_{-39}$ & $0.131^{+32}_{-32}$ & \phantom{$-$}$0.017^{+43}_{-43}$ & $2800^{+680}_{-680}$ & \phantom{$-$}$360^{+920}_{-920}$ & $0.157^{+0.038}_{-0.038}$ & 640 & 0.94 \\
NNE & 72 & $0.038^{+13}_{-13}$ & \phantom{$-$}$0.104^{+14}_{-13}$ & $0.111^{+12}_{-12}$ & $-0.010^{+14}_{-14}$ & $2360^{+260}_{-260}$ & $-200^{+300}_{-300}$ & $0.154^{+0.017}_{-0.017}$ & 650 & 0.92 \\
SE1 & 69 & $0.089^{+18}_{-17}$ & \phantom{$-$}$0.023^{+24}_{-23}$ & $0.071^{+20}_{-19}$ & $-0.058^{+23}_{-23}$ & $1520^{+420}_{-410}$ & $-1240^{+490}_{-490}$ & $0.103^{+0.028}_{-0.028}$ & 970 & 0.62 \\
SE2 & 86 & $0.033^{+18}_{-18}$ & $-0.084^{+27}_{-29}$ & $0.075^{+23}_{-23}$ & \phantom{$-$}$0.049^{+24}_{-23}$ & $1610^{+490}_{-490}$ & \phantom{$-$}$1040^{+510}_{-480}$ & $0.088^{+0.027}_{-0.027}$ & 1140 & 0.53 \\
SSE & 71 & $0.039^{+12}_{-13}$ & $-0.058^{+14}_{-14}$ & $0.070^{+14}_{-14}$ & $-0.002^{+13}_{-13}$ & $1490^{+300}_{-300}$ & $-50^{+270}_{-280}$ & $0.099^{+0.020}_{-0.020}$ & 1010 & 0.59 \\
\enddata
\tablecomments{PM errors (in parentheses) are in milliarcsec yr$^{-1}$.}
\tablenotetext{a}{R.A.~PM.}
\tablenotetext{b}{Decl.~PM.}
\tablenotetext{c}{Radial PM (away from the CCO).}
\tablenotetext{d}{Tangential PM, negative (positive) for clockwise (counterclockwise) motion.}
\tablenotetext{e}{Velocities assume a distance of 4.5 kpc.}
\tablenotetext{f}{Free expansion age (distance over $v_r$).}
\tablenotetext{g}{Relative deceleration $m \equiv vt/R$, where the
  age $t$ is assumed to be 600 yr, the actual upper limit.  So
  $m$ values are relative to region B3, the region with the 
shortest expansion age.}
\end{deluxetable*}

Figure~\ref{motions} and Table~\ref{rates} present the basic
results of our investigation.  The expansion away from the CCO is
dramatic and obvious.  The highest velocities occur at the eastern
edge, above the indentation, where motions appear to diverge around an
unseen obstacle.  Fainter regions to the southeast have markedly
nonradial motions as well.  Considerable variations in speed can be
seen within the bright eastern area (top panel of
Figure~\ref{motions}).

The remnant is of course younger than the shortest expansion age we
find in Table~\ref{rates}, about 600 yr, independent of
distance.  We conclude that \src\ is one of the three youngest known
CC SNRs in the Galaxy, no more than twice the age of Cas A, and
comparable to the age of Kes 75.  We assume a nominal age
of 600 yr for estimates below.

Table~\ref{rates} also includes relative decelerations $m$, that
is, ratios of the nominal 600 yr age to the expansion ages.  Since
the true age may be less than 600 yr, the decelerations are upper
limits.  A wide range is apparent.

\section{Spectroscopy}

We examined spectra of several small regions that stood out
kinematically.  The most rapid expansion is found in regions A1--A3,
with an average radial velocity (away from the neutron star) of
about $5100 d_{4.5}$ km s$^{-1}$.  Substantial deceleration is
apparent in regions C1 and C2, which expand at about $3000 d_{4.5}$ km
s$^{-1}$.  The bright filament slightly further west 
(within regions E1--E4) expands slightly faster (about
($3200$--$4000$) $d_{4.5}$ km s$^{-1}$), but has contrasting spectral properties.
A much fainter region well separated from the bright eastern area is
region NNE, expanding at only about $2300 d_{4.5}$ km s$^{-1}$ (but
relatively undecelerated, i.e., high $m$).

\begin{figure}
  \centerline{\includegraphics[width=3.5truein]{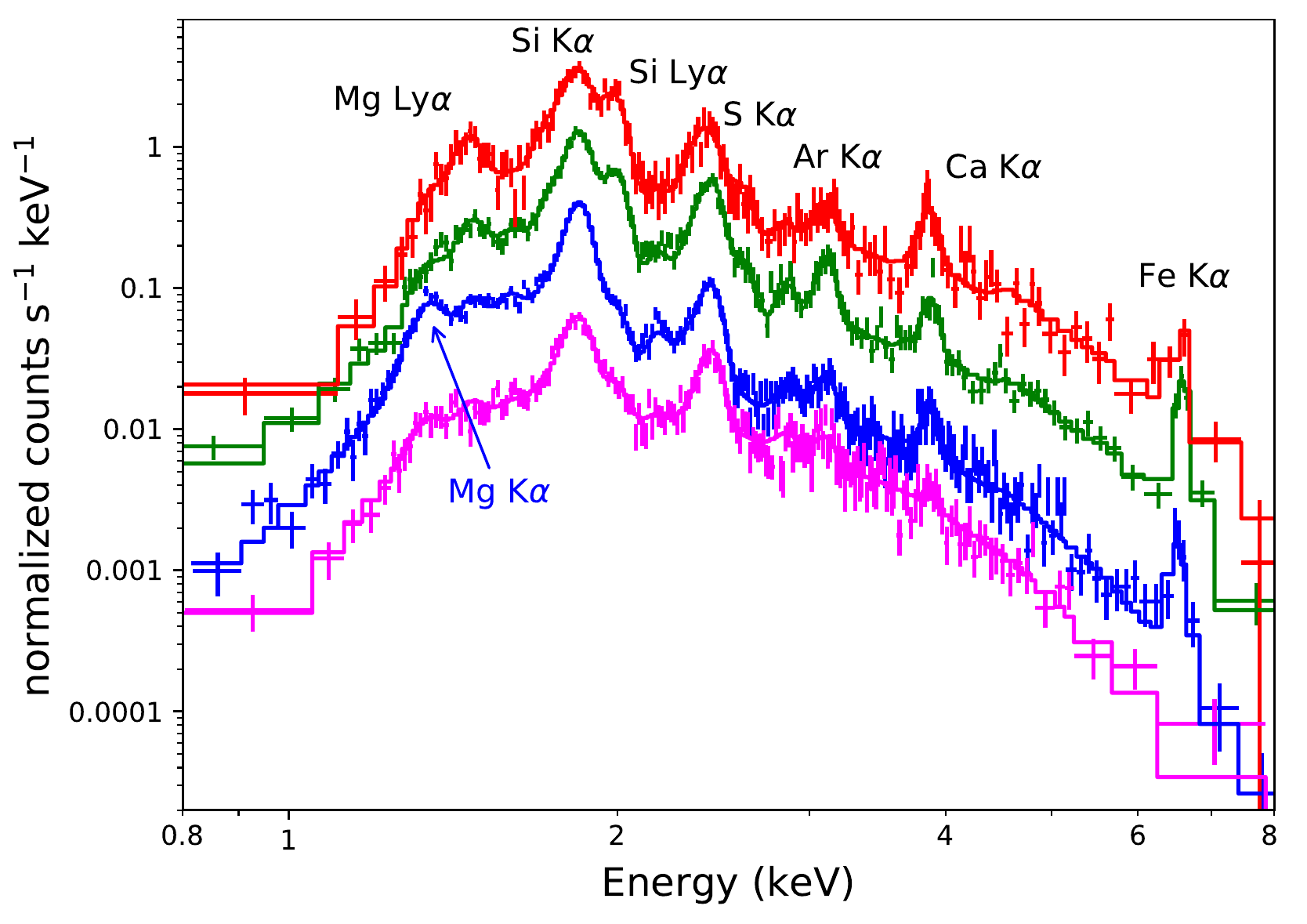}}
  \caption{Spectra of four regions, with the most prominent spectral features
    labeled.
    From the top down, shifted upward by 2, 1, 0.25, and 0 dex:  regions A1--A3 
    combined, regions C1 and C2, bright filament within region set E, and
    region
    NNE. The first three move from E to W through the bright area. Note
    the drop in ionization state, from the decreasing prominence of
    Si Ly$\alpha$ at 2 keV.  Note also the strong presence of Fe K$\alpha$ in
    all but region NNE.}
  \label{spectra}
  \end{figure}
  
The spectra are shown in Figure~\ref{spectra}.  We fit them with
plane-shock models with absorption that are available in
Xspec \citep{arnaud96}, using the
abundance set from \cite{grsa98}.  For fainter regions A1--A3 and NNE,
backgrounds were modeled rather than subtracted, allowing the use of
C-statistics \citep{cash79}.
Region NNE is well described by a plasma with subsolar ($0.75
(0.63, 0.92)$) abundances 
(with solar heavy-element abundance ratios assumed),
with temperature $kT = 1.00\, (0.94, 1.09)$ keV,
ionization age $\tau = 4.0\, (3.0, 5.6) \times 10^{11}$ cm$^{-3}$ s,
a redshift of $900\, (800, 1300)$ km s$^{-1}$, and hydrogen column
$N_{\rm H} = 4.2\, (4.0, 4.4) \times 10^{22}$ cm$^{-2}$ (errors
are 90\%\ confidence intervals). We conclude
that it is dominated by shocked circumstellar or interstellar material
(CSM/ISM); the apparently subsolar abundances may simply indicate that
an additional continuum component, perhaps nonthermal, is also
present, in addition to that due to H and He.  The other three regions
all require extremely oversolar abundances, suggesting that both lines
and continua originate in heavy-element ejecta.
Since O and heavier elements are primary nucleosynthetic products
of CC SNe, we set abundances of elements lighter
than O to zero, and fit for abundances
of Mg, Al, Si, S, Ar, Ca, and Fe (with the relative IGE abundances
fixed to solar ratios, and with Ne and odd-$Z$ elements other than Al
set to solar values).
There is
Fe present in all three spectra, with $[{\rm Fe}/{\rm O}] = 0.28\,
(0.18, 0.68)$ (regions A1--A3), $0.64\, (0.41, 1.06)$ (C1 and C2), and
$2.1\, (0.9, 4.1)$ (region E). For Si and S, the intermediate-mass
elements with the most prominent lines, $[{\rm Si}/{\rm O}] = 1.3\,
(1.0, 2.2)$, $1.3\, (1.1, 1.6)$, $1.5 \, (1.1, 2.6)$, and $[{\rm S}/{\rm
    O}] = 0.42\, (0.31, 0.67)$, $0.63 \, (0.54, 0.77)$, and $0.60\, (0.44,
0.86)$, respectively.
Progressing from east to west, those three spectra show
systematically decreasing temperature from $2.6\, (2.0, 3.4)$ keV,
through $2.2\, (1.9, 2.6)$ keV to $1.3\, (1.2, 1.4)$ keV, and also
decreasing $\tau$ from $3.4\, (2.3, 5.3) \times 10^{11}$ cm$^{-3}$ s, 
through  $2.5\, (2.0, 3.3) \times 10^{11}$ cm$^{-3}$ s to
$1.7 (1.4, 2.1)\, \times 10^{11}$ cm$^{-3}$ s. 
We find considerable line broadening and Doppler shifts for all three regions:
all are redshifted, with
A1--A3 moving away at $2600\, (2500, 2700)$ km s$^{-1}$.
With our measured PM, we
obtain a space velocity of 5800 km s$^{-1}$ at an angle of 27$^\circ$
with the plane of the sky. The other two regions in the bright area
also show recession speeds, with C1 and C2 at 
$1300\, (900, 1600)$ km s$^{-1}$, and region set E at $1100\, (800, 1300)$
km s$^{-1}$. Regions A1--A3 show the most extreme broadening,
with ($1\sigma = {\rm FWHM}/2.35$) line widths of $2500\, (2400, 4500)$ km s$^{-1}$.

\section{Results and Discussion}

Here we summarize our results.

\begin{enumerate}

\item We observe radial expansion away from the compact central object
  (CCO) in all directions in \src, at speeds ranging up to $5800\,
  d_{4.5}$ km s$^{-1}$.  The remnant age is less than the shortest
  expansion time we measure, about 600 years.  We conclude that
  \src\ is the remnant of one of the three most recent known
  core-collapse supernovae in the Galaxy.

\item The CCO might be moving in the NNW direction with $v_t \sim 300 d_{4.5}$ km s$^{-1}$,
  a rather typical neutron star speed,
  but this is quite uncertain because of large PM errors. With this speed,
  the CCO must be still quite near 
  the explosion center (within $10''$).

\item Nonradial motions are apparent in several regions.  In the
  bright east quadrant, we observe flow around some X-ray-dark obstacle.
  
\item We find that pure heavy-element ejecta devoid of H and He 
  can explain all the emission seen 
  in almost all the
  bright regions of \src, indicating that the shocked ejecta
  strongly dominate over any CSM/ISM emission there. In addition to
  intermediate-mass elements that dominate the X-ray spectra, there is
  freshly synthesized iron within these fast-moving ejecta. 
  Some iron-rich regions are expanding at 3000--5000 km s$^{-1}$.
  The region with the fastest expansion PMs also shows a
  redshift of 2000--3000 km s$^{-1}$, for a space velocity of almost
  $6000 d_{4.5}$ km s$^{-1}$.

\item We find markedly different conditions in faint parts of \src,
    in one of which we find no evidence for supersolar abundances.

\end{enumerate}
    
The extreme spatial asymmetry of the X-ray emission with respect to
the CCO must be due to some combination of intrinsic asymmetry in the
explosion and asymmetric surroundings.  Evidence of strong asymmetry
in the explosion itself is provided by our finding of iron
within very high velocity SN ejecta. 
Rayleigh-Taylor instabilities in expanding
supernova ejecta have been invoked to explain elemental asymmetries
inferred from observations of supernovae \citep[e.g.,][]{orlando16},
with 3D simulations showing ``fingers'' of iron-group elements
penetrating to larger radii and velocities \citep{gabler20}.  However,
the range of models shown there failed to produce significant masses
of iron moving at 4000 km s$^{-1}$ or faster.  A study of SN 2013ej, a
peculiar SN IIP \citep{utrobin17}, invoked a jetlike explosion in a
red supergiant to fit the light curve, and deduced mixing of IGEs to
4000--6500 km s$^{-1}$.

Evidence for a highly irregular CSM is also strong.  A low velocity
for the CCO argues against the east region representing the bulk of the
ejecta, as it suggests equal momentum ejected in the opposite direction.
The absence of bright emission to the west then indicates
lower densities in that direction. Clearly absent is any strong indication of
CCO motion opposite to the bulk of observed X-rays as found in five of six
more symmetric remnants studied by \cite{holland17}. 
At a smaller scale, the interaction of ejecta
with some dark obstacle in the east is very obvious, requiring a substantially
higher density there.  While the deceleration caused by this obstacle can be
partly responsible for the much greater brightness in this area, the
large deceleration we measure in southeast regions, which are quite faint,
rules out any simple relation between deceleration and brightness.  We
conclude that both the ejecta and surrounding material in \src\ are
highly inhomogeneous.

We located one faint region (NNE) that appears to be relatively free
of ejecta, suggesting that it represents a part of the blast wave
interacting with circumstellar or interstellar material.  However, we
were unable to trace any coherent features corresponding to the blast
wave on larger scales.  Very blue (hard-spectrum) emission region H1
(see Figures~\ref{3colors} and~\ref{regions}) may be nonthermal
emission, but scattered light from the adjacent bright emission makes
analysis difficult.

The brightest part of \src\ is made up of ejecta heated in a reverse
(inward-facing) shock.  Material farther from the CCO would have been
shocked longer ago, so one might expect systematically increasing
ionization ages as one moves east.  Furthermore, if electrons are
progressively heated by Coulomb collisions rather than in some
collisionless process at the shock, the electron temperature should
also increase in that direction.  We observe both trends (see
Figure~\ref{spectra} and the corresponding text).  However, the spatial
separation of those regions is almost certainly too large to be due to
this effect in a single well-defined shock.  The absence of an obvious
feature corresponding to a reverse shock strengthens this conclusion.
It is more likely that projection effects and multiple shocks are
involved.

High-velocity Fe ejecta can be produced more easily in asymmetric
models.  \cite{utrobin17} showed that observations of several SN IIP
events (i.e., from progenitors with extended envelopes)
require high-velocity Fe, and modeled them successfully with
highly asymmetric (bipolar) ejection of IGEs.
On the other hand, Cas A resulted from a mostly stripped
SN IIb progenitor, while SN 1987A's progenitor was a blue
supergiant---i.e., both events without extended envelopes.  SN ejecta are
expected to be least decelerated on average for SNe IIb, Ib, and Ic
because of their relatively low ejecta masses.  Maximum IGE
velocities over 5000 km s$^{-1}$ might be common in
these stripped-envelope SNe.  However, models of
artificially mixed SNe Ibc \citep{woosley20} are subluminous compared
to observations, and may be subenergetic as well.  
So for \src, there is a slight preference
for a stripped-envelope event as opposed to an SN IIP. But with no
data from the supernova event that created \src, we cannot constrain
the nature of the progenitor further.

The high Ni overabundance found by \citet{yasumi14}, with a Ni/Fe mass
ratio of $0.7 \pm 0.4$ ($12 \pm 7$ times solar), while uncertain, is
quite intriguing in view of the very high velocity Fe that we found in
\src. Such elevated abundances of Ni are quite rare;
\cite{jerkstrand15a} reported Ni/Fe $= 3.4 \pm 1.2$ times solar in SN
2012ec, and collected observations of several other SNe for which this
determination is possible. \cite{jerkstrand15b} used parameterized
thermodynamic trajectories to constrain the conditions for high Ni/Fe
production, and showed that spherically symmetric models with lower
progenitor mass are able to accomplish this.  However, such 1D models
do not explode.  The asymmetries required for successful CC supernova
explosions in simulations may also eject more neutron-rich IGEs for a
wider mass range of progenitors.  If future observations confirm the
high Ni/Fe ratio in \src, it is possible that the explosion
asymmetries required for the fast Fe may also play a role in stable
nickel production.  

Finally, we note that a distance as large as 9 kpc \citep{yasumi14} 
would imply 12,000 km s$^{-1}$ ejecta velocities, with some
$^{56}$Ni ejected with such astonishing speeds during the explosion.
However, the high column densities 
($N_{\rm H} \sim 4.2 \times 10^{22}$ cm$^{-2}$) argue against a distance much
less than 4.5 kpc.

\section{Conclusions}

Our observations indicate that \src\ is an extremely asymmetric
remnant, with an expansion center at or near the CCO and expansion
velocities of up to $6000 d_{4.5}$ km s$^{-1}$.  These velocities give expansion
ages of as little as 600 yr, so that the remnant is younger than
this---the third youngest known CC SNR in the Galaxy.  The presence of
ejecta, iron in particular, at such high velocities is strong evidence
in favor of \src\ having resulted from an asymmetric, perhaps
stripped-envelope supernova event. As such, it represents the most
extreme case among remnants of this phenomenon, and warrants careful
study.  More detailed spectroscopic analysis is
possible with these data, and can cast light on this phenomenon.

\acknowledgments We gratefully acknowledge support by NASA through
                 {Chandra} General Observer Program grant SAO
                 GO8-19053X.

\vspace{5mm}
\facilities{CXO}


\begin{thebibliography}{}

\bibitem[Arnaud(1996)]{arnaud96}
  Arnaud, K. A. 1996, in ASP Conf.~Ser.101, Astronomical Data Analysis and
  Systems V, ed.~G.~Jacoby \& J.~Barnes (San Francisco, CA: ASP), 17

\bibitem[Borkowski et al.(2020)]{borkowski20}
  Borkowski, K.~J., Reynolds, S.~P., \& Miltich, W. 
  2020, ApJL, 895, L32

\bibitem[Borkowski et al.(2018)]{borkowski18}
  Borkowski, K.~J., Reynolds, S.~P., Williams, B.~J., \& Petre, R.
  2018, ApJL, 868, L21

\bibitem[Cash(1979)]{cash79}
  Cash, W. 1979, ApJ, 228, 939

\bibitem[DeLaney et al.(2010)]{delaney10}
  DeLaney, T., Rudnick, L., Stage, M.~D., et al.
  2010, ApJ, 725, 2038

\bibitem[Gabler et al.(2020)]{gabler20}
  {Gabler}, M., Wongwathanarat, A., \& Janka, H.-T.
  2020, arXiv:2008.01763

\bibitem[Gaensler et al.(2008)]{gaensler08}
  Gaensler, B. M., Tanna, A., Slane, P. O., et al.
  2008, ApJL, 680, L37
  
\bibitem[Grevesse \& Sauval(1998)]{grsa98}
Grevesse, N., \& Sauval, A. J.
1998, SSRv, 85, 161

\bibitem[Holland-Ashford et al.(2017)]{holland17}
Holland-Ashford, T., Lopez, L. A.,  Auchettl, K., Temim, T., \& Ramirez-Ruiz, E.
2017, ApJ, 844, 84 

\bibitem[Houck \& Fransson(1996)]{houck96}
  Houck, J. C., \& Fransson, C.
  1996, ApJ, 456, 811

\bibitem[Jerkstrand et al.(2015a)]{jerkstrand15a}
  Jerkstrand, A., Smartt, S. J., Sollerman, J., et al.
  2015a, MNRAS, 448, 2482

\bibitem[Jerkstrand et al.(2015b)]{jerkstrand15b}
  Jerkstrand, A., Timmes, F. X., Magkotsios, G., et al.
  2015b, ApJ, 807, 110

\bibitem[Lovchinsky et al.(2011)]{lovchinsky11}
  Lovchinsky, I., Slane, P., Gaensler, B.~M., et al.
  2011, ApJ, 731, 70

\bibitem[McCray \& Fransson(2016)]{mccray16}
  McCray, R., \& Fransson, C.
  2016, ARAA, 54, 19

\bibitem[Orlando et al.(2016)]{orlando16}
  Orlando, S., Miceli, M., Pumo, M. L., \& Bocchino, F. 
  2016, ApJ, 822, 22
  
\bibitem[Orlando et al.(2020)]{orlando20}
  Orlando, S., Ono, M., Nagataki, S., et al. 
  2020, A\&A, 636, A22

\bibitem[Reynolds et al.(2018)]{reynolds18}
  Reynolds, S. P., Borkowski, K. J., \& Gwynne, P.~H.
  2018, ApJ, 856, 133

\bibitem[Salter et al.(1986)]{salter86}
  Salter, C. J., Patnaik, A. R., Shaver, P. A., \& Hunt, G. C.
  1986, A\&A, 162, 217
  
\bibitem[Thorstensen et al.(2001)]{thor01}
  Thorstensen, J.~R., Fesen, R.~A., \& van den Bergh, S.
  2001, AJ, 122, 297

\bibitem[Utrobin \& Chugai(2017)]{utrobin17}
  Utrobin, V. P., \& Chugai, N. N.
  2017, MNRAS, 472, 5004

\bibitem[Utrobin \& Chugai(2019)]{utrobin19}
  Utrobin, V. P., \& Chugai, N. N.
  2019, MNRAS, 490, 2042

\bibitem[Wongwathanarat et al.(2017)]{wongwathanarat17}
  Wongwathanarat, A., Janka, H.-T., M{\"u}ller, E., Pllumbi, E., \& Wanajo, S.
  2017, ApJ, 842, 13

\bibitem[Woosley et al.(2020)]{woosley20}
  Woosley, S., Sukhbold, T., \& Kasen, D.
  2020, arXiv:2009.06868
  
\bibitem[Yasumi et al.(2014)]{yasumi14}
  Yasumi, M., Nobukawa, M., Nakashima, S., et al.
  2014, PASJ, 66, 68

\end{thebibliography}
\end{document}